\documentclass[twocolumn,prb,floatfix,showpacs]{revtex4}

\usepackage{amsmath,array,dcolumn,bm}

\usepackage{epsfig}

\usepackage{times,mathptmx}

\begin{document}

\title{Structure and formation energy of carbon nanotube caps}

\author{S. Reich}
\affiliation{Department of Engineering, University of Cambridge, Trumpington Street, Cambridge CB2 1PZ, United Kingdom}

\author{L. Li}
\affiliation{Department of Engineering, University of Cambridge, Trumpington Street, Cambridge CB2 1PZ, United Kingdom}

\author{J. Robertson}
\affiliation{Department of Engineering, University of Cambridge, Trumpington Street, Cambridge CB2 1PZ, United Kingdom}

\begin{abstract}
We present a detailed study of the geometry, structure and energetics of carbon nanotube caps. We show that the structure of a cap uniquely determines the chirality of the nanotube that can be attached to it. The structure of the cap is specified in a geometrical way by defining the position of six pentagons on a hexagonal lattice. Moving one (or more) pentagons systematically creates caps for other nanotube chiralities. For the example of the (10,0) tube we study the formation energy of different nanotube caps using \emph{ab-initio} calculations. The caps with isolated pentagons have an average formation energy $(0.29\pm0.01)$\,eV/atom. A pair of adjacent pentagons requires a much larger formation energy of 1.5\,eV. We show that the formation energy of adjacent pentagon pairs explains the diameter distribution in small-diameter nanotube samples grown by chemical vapor deposition.\end{abstract}

\pacs{61.46.+w,61.48.+c,81.10.Aj}

\maketitle

\section{Introduction}

When carbon nanotubes were discovered more than ten years ago, they were generally looked at as elongated fullerenes.\cite{iijima91,iijima93,dresselhausbuch,dresselhaus92,saito92,yu95} A tube was described as several carbon rings capped on both ends by half-spherical fullerene fragments. Over the years this view changed. Presently, carbon nanotubes are modeled as one-dimensional solids.\cite{reichbuch} A cylindrical unit cell is repeated infinitely along the nanotube axis.\cite{reichbuch} Considering single-walled carbon nanotubes as systems with translational periodicity is appropriate for studying their physical properties, because nanotube aspect ratios range from $10^2 - 10^7$ (Refs.\onlinecite{hata04} and \onlinecite{zheng04}). The caps---normally only present at one end of the tubes---have virtually no effect onto the properties of a tube. This is the reason why nanotube caps almost disappeared from the literature, after a number of studies on this subject in the earlier nanotube literature.\cite{fujita92,astakhova99,brinkmann99} 

Recently, however, the attention turned back to nanotube caps in efforts to understand the growth of carbon nanotubes. Single-walled tubes nucleate on a catalyst particle and then normally grow by adding carbon atoms to the base (root growth mechanism).\cite{saito94,gavillet01,fan03,hata04} Miyauchi~\emph{et al.}\cite{miyauchi04} pointed out the importance of caps when studying small-diameter tubes grown by chemical vapor deposition (CVD). They argued that there must be a correlation between the apparent preference for some nanotube chiralities in their growth method and the structures of the nanotube caps at nucleation. We recently suggested that the caps of carbon nanotubes and their interaction with the catalyst will be the key for controlling the chirality of single-walled carbon nanotubes during growth.\cite{reich05c} Our guiding principle for chirality-selective growth was based on two fundamental concepts: (i) one cap can only grow into a unique carbon nanotube and (ii) the cap structure can be controlled by epitaxial growth on a metal surface.

In this paper we study the correlation between nanotube caps and the tube that can be attached to it. We show that the matching nanotube structure is defined by placing six pentagons onto a hexagonal lattice. Moving one of the pentagons creates different caps and hence different tubes in a rational way. Moving one pentagon along the graphene $\bm{a}_1$ direction changes the nanotube chirality $(n_1,n_2)$ to $(n_1,n_2+1)$, moving it along $\bm{a}_2$ creates the $(n_1-1,n_2+1)$ tube. We study the energetics of various (10,0) nanotube caps using \emph{ab-initio} calculations. For caps obeying the isolated pentagon rule the average formation energy is $17.4\pm0.7\,$eV per 60 carbon atoms. One pair of adjacent pentagons requires an additional energy of 1.5\,eV. We show that the large formation energy of adjacent pentagons explains the diameter distribution in low-temperature CVD grown samples reported by Bachilo~\emph{et al.}\cite{bachilo03} and Miyauchi~\emph{et al.}\cite{miyauchi04}.

This paper is organized as follows. We first show how to construct a nanotube cap by cutting out 60$^\circ$ cones from a hexagonal lattice, Sect.~\ref{sec_construction}. We discuss the number of possible caps for a given nanotube---several hundreds to thousands for tube diameters $d\approx10\,${\AA}---and the number of possible nanotubes for a given cap---just one in Sects.~\ref{sec_construction} and \ref{sec_one_cap}. We then show in Sect.~\ref{sec_different_tubes} how caps for different tubes can be constructed by moving pentagons. After explaining the construction of nanotube caps we proceed by calculating the formation energies of a total of 20 cap structures for the (10,0) nanotube in Sect.~\ref{sec_formation}. Finally, we discuss the diameter distribution of small-diameter tubes ($d=7-10\,${\AA}) grown by CVD, Sect.~\ref{sec_abundance}. Section~\ref{sec_conclusion} summarizes this work.

\section{Cap construction}\label{sec_construction}

\begin{figure}[b]
\epsfig{file=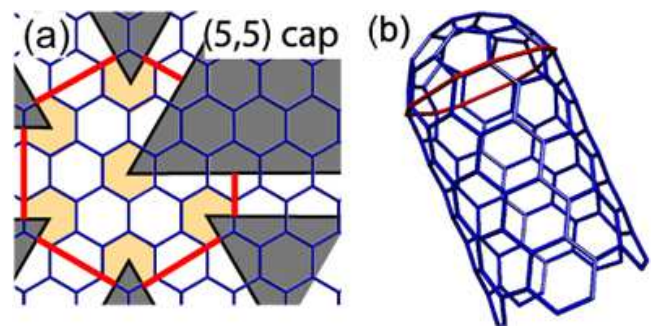,width=8.5cm,clip=}
\caption[]{(Color online) Constructing a cap for a (5,5) nanotube. (a) Cut out the shaded areas and join the sides to obtain the cap in (b). Light grey (orange) mark the pentagons. The cap hexagon (thick line, red) in (a) points around the nanotube circumference in (b).}
\label{cap_55}
\end{figure}

Carbon caps resemble half-fullerenes. They are composed of six pentagons and a number hexagons.\cite{brinkmann99,astakhova99,fujita92} 
The six pentagons are necessary by Euler's theorem
of closed polyhedra to introduce the necessary Gaussian curvature.\cite{dresselhausbuch,terrones92,terrones93,lenosky92}
There are three methods to represent carbon caps on a flat plane: flattening the cap onto a hexagonal lattice\cite{yoshida95}, unwrapping a half tube with the cap attached to it\cite{fujita92}, and a network representation based on graph theory\cite{brinkmann99}. We use the flattening method originally suggested by Yoshida and Osawa\cite{yoshida95}, see also Astakhova~\emph{et al.}\cite{astakhova99}. We found that this method best highlights the pattern of six hexagons and its correlation to the nanotube chiral vector.

Figure~\ref{cap_55}(a) shows the construction of a (5,5) nanotube cap using the flattening method. The light grey (orange) hexagons indicate the positions of the pentagons on the hexagonal graphene lattice. The dark shaded areas are cut and the black lines joined to form a half-spherical structure,\cite{astakhova99} see Fig.~\ref{cap_55}(b). The (red) lines in Fig.~\ref{cap_55}(a)---the cap hexagon---define the rim of the cap; this line goes around the circumference in the capped (5,5) in (b).

The (5,5) nanotube is the smallest diameter tube that has a cap obeying the isolated pentagon rule.\cite{brinkmann99} It has only a single such structure fitting over it (half a C$_{60}$ fullerene). The other tubes with only one cap fulfilling the isolated pentagon rule are (9,0), (9,1), (8,2), and (6,5) (diameters $d=6.8-7.5\,${\AA}). Tubes of these diameters are found, \emph{e.g.}, in CoMoCat CVD grown samples.\cite{bachilo03,miyauchi04} With this growth method even smaller tubes such as the (6,4) and the (7,3) have been reported. Allowing adjacent pentagons the smallest capped tube is the (5,0) tube. This sets the limit for the tube diameters that can form by the standard growth method. Interestingly, for the small-diameter CVD samples there is a marked decrease in the abundance of tubes when going from the isolated pentagon regime [(6,5), (8,3), (9,1)] to the diameters where only caps with adjacent pentagons exist [(6,4) and (7,3)]. We discuss the diameter distribution in these samples in Sect.~\ref{sec_abundance}.

\begin{figure}[t]
\epsfig{file=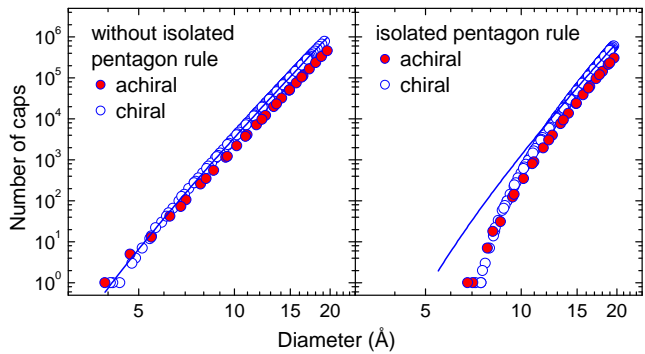,width=8.5cm,clip=}
\caption[]{(Color online) Number of carbon caps \emph{versus} tube diameter; (a) all possible cap structures and (b) cap structures obeying the isolated pentagon rule. The line in (a) is a fit by a function $\lvert d-d_c\rvert^v$ with $v=7.8$ and $d_c=1.2$; in (b) $d_c=2.6$. Data from Ref.~\onlinecite{brinkmann99}.}
\label{number}
\end{figure}

Brinkmann~\emph{et al.}\cite{brinkmann99} calculated the number of caps for nanotube diameters up to 20\,{\AA}. Similar numbers as reported by Brinkmann and co-workers were obtained by Astakhova~\emph{et al.}\cite{astakhova99} for the subset of caps obeying the isolated pentagon rule and tubes up to the (10,10) nanotube. 
With increasing diameter the number of possible caps grows very rapidly following a power-law for caps with two or more adjacent pentagons, see Fig.~\ref{number}(a) (Ref.~\onlinecite{brinkmann99}). 
In a first approximation one expects a $d^{12}$ dependence for the number of caps (including adjacent pentagons) by the following argument. The area of the hexagonal lattice where we can place the six pentagons as in Fig.~\ref{cap_55} grows as $d^2$, because the cap hexagon is equal to the circumference of the tube. To specify the cap by the flattening method we select six out of $d^2\approx h^2$ elements ($h$ integer). We show in Sect.~\ref{sec_one_cap} that specifying the six pentagons fully determines the chirality of the nanotube. The number of distinct choices for six elements out of $h^2$ is given by the binomial coefficient $_{h^2}C_5$. The highest power of this binomial is $h^{12}$; from this we obtain the $d^{12}$ dependence for the number of carbon caps. As can be seen from the fit in Fig.~\ref{number}(a) the number of caps is proportional to $d^{7.8}$. There are three obvious reasons why the number of caps is smaller for a given $d$ than expected from our argument.

First, the choice of the six pentagons is not independent. By placing the first pentagon we remove a 60$^\circ$ cone from the hexagonal lattice, Fig.~\ref{cap_55}(a). The remaining pentagons must not be placed in this area. This becomes more and more restrictive the more pentagons we specify. Second, two apparently different sets of six pentagons out of $h^2$ enumerated elements can, in fact, correspond to the same geometrical pentagon pattern. The pattern might be rotated or moved with respect to the hexagonal lattice. For example, in Fig.~\ref{cap_55}(a) we can rotate the six pentagons by 60$^\circ$ degree around the central pentagon (the cones are rotated together with the pentagons). This is a different subset in the argument outlined above, but the two caps are identical. Third, the cap area is not necessarily quadratic; it can be rectangular as well [examples are the (8,4) caps in Fig.~\ref{different_cuts} of Sect.~\ref{sec_one_cap}]. Then, the number of available hexagons is smaller than $h^2$, which results in a smaller binomial coefficient and hence a smaller exponent for the number of caps over tube diameter. Although we have several arguments why the exponent in Fig.~\ref{number}(a) is smaller than twelve, we cannot explain why it is equal to $7.8$. It would be interesting to explain the exact dependence of the number of caps on tube diameter.

For caps fulfilling the isolated pentagon rule in Fig.~\ref{number}(b) the number of caps is smaller for small diameters than for general caps, but for $d\rightarrow\infty$ we recover the power law behavior of Fig.~\ref{number}(a).\cite{brinkmann99} This is understandable, since for a large cap area (large tube diameters) the fraction of caps with adjacent pentagons becomes negligible.
For a given nanotube diameter there are fewer caps for armchair and zigzag (closed symbols) tubes than for chiral tubes (open). This is due to the higher symmetry of the achiral tubes, which reduces the choices of caps. 

Although a given nanotube can have thousands of distinct caps, quite the opposite is true for the inverse problem. A given cap only fits onto one particular nanotube as we show in the following section. 

\subsection{One cap -- one tube: The pentagon pattern determines the tube chirality}\label{sec_one_cap}

\begin{figure}[b]
\epsfig{file=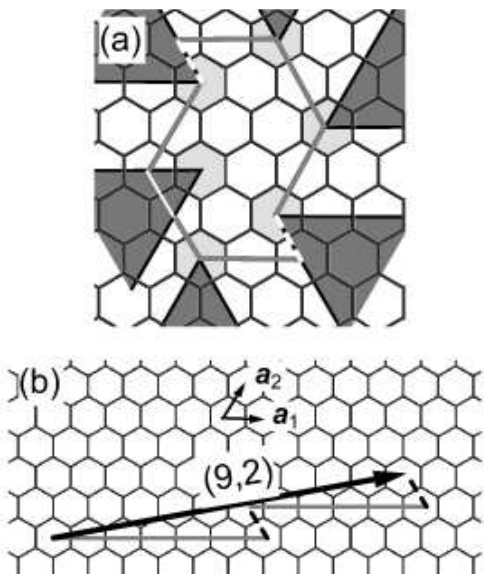,width=8.5cm,clip=}
\caption[]{(Color online) (a) pentagon pattern, cutting direction and cap hexagon of a (9,2) cap. (b) The sum of the full (red) and dashed (white) lines in (a) equals the (9,2) chiral vector (arrow).}
\label{how_to}
\end{figure}

As we showed in Fig.~\ref{cap_55}(a) and (b) a hexagon that includes all six pentagons on the graphene lattice ends up as a closed line around the nanotube circumference. Giving the vector around the nanotube circumference, on the other hand, uniquely determines the microscopic structure of the tube (diameter $d$ and chiral angle $\varTheta$). As is well known an $(n_1,n_2)$ nanotube is obtained by rolling up a graphene sheet along the vector\cite{reichbuch}
\begin{equation}
\bm{c}=n_1\bm{a}_1+n_2\bm{a}_2,
\end{equation}
where $\bm{a}_1$ and $\bm{a}_2$ are the graphene lattice vectors. Once the cap hexagon is defined as in Fig.~\ref{cap_55}(a) the edge of this cap fixes the nanotube structure that can be attached to it. 

In Fig.~\ref{how_to} we show the edge construction for a chiral nanotube. In this case the cap hexagon cannot be closed by cutting the shaded areas. The steps [dashed lines in Fig.~\ref{how_to}(a)] correspond to going twice into the $-\bm{a}_1+\bm{a}_2$ direction of graphene, see Fig.~\ref{how_to}(b). The full (red) line representing the cap hexagon has a length of $11\bm{a}_1$. Adding the full and dashed lines we obtain a chiral vector $\bm{c}=(11-2)\bm{a}_1+2\bm{a}_2$, \emph{i.e.}, the (9,2) nanotube as shown in Fig.~\ref{how_to}(b).

\begin{figure}[t]
\epsfig{file=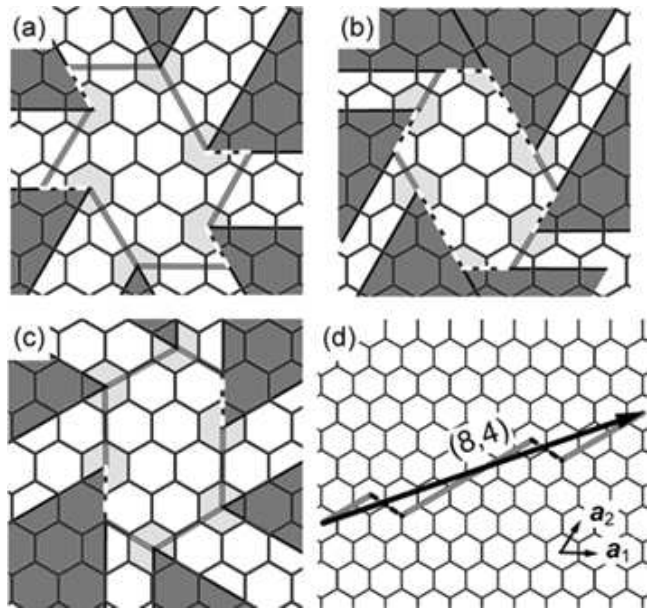,width=8.5cm,clip=}
\caption[]{(Color online) (a) Construction of an (8,4)  cap. (b) The same arrangement of the six pentagons but a rotation of the cut cones (dark grey areas) yields the same tube, see text. (c) Cutting along the armchair direction and rotating the cap hexagon yields likewise an (8,4) cap as shown in (d).}
\label{different_cuts}
\end{figure}

The cap construction by flattening the half sphere onto the graphene lattice involves three steps:\cite{astakhova99} (i) define the positions of the hexagons, (ii) define the directions of the $60^\circ$ cones to be cut and (iii) define the cap hexagon. Which step fixes the nanotube structure? 

The last step---drawing the cap hexagon---is completely arbitrarily and will not change the cap structure. Increasing the size of the hexagon by moving the lines in Fig.~\ref{how_to}(a) away from the pentagons, will simply add a tube segment to the cap. It is also possible to ``rotate'' the hexagon, \emph{i.e.}, to change the orientation of the line connecting two pentagons by 60$^\circ$ and correspondingly the remaining lines between the other pentagons. This changes the orientation of the full lines and the steps with respect to $\bm{a}_1$ and $\bm{a}_2$. Adding up the lines and steps one obtains the same chirality.

How abou step (ii) of the flattening method? Can we construct two distinct caps from a given arrangement of pentagons by cutting different segments of the graphene lattice? The answer is no as is illustrated for the (8,4) nanotube in Fig.~\ref{different_cuts}(a)-(d). Cutting different segments of the graphene lattice [Fig.~\ref{different_cuts}(a) and (b)] results in the same chiral vector. The construction of the (8,4) cap in Fig.~\ref{different_cuts}(a) is equivalent to the construction of the (9,2) cap in Fig.~\ref{how_to}(a). The full line has a length of 12$\bm{a}_1$ and we have four steps along $-\bm{a}_1+\bm{a}_2$. This adds up to the (8,4) nanotube. In Fig.~\ref{different_cuts}(b) we rotated the shaded cones by 60$^\circ$ degree. On first sight the cap looks remarkably different. The full lines have a length of $4\bm{a}_1$ (we rotated the cap hexagon together with the shaded cones to keep the full lines parallel to $\bm{a}_1$). There are a total of 8 steps; they are along $\bm{a}_1-\bm{a}_2$. Thus, Fig.~\ref{different_cuts}(b) shows a cap for the $(12,-4)$ tube. This is symmetry equivalent to the (8,4) nanotube because the $(n_1,n_2)$ and the $(n_1+n_2,-n_2)$ tube have the same microscopic structure.\cite{reichbuch} In Fig.~\ref{different_cuts}(c) we cut along the armchair instead of the zigzag direction. Once more, we obtain a cap for the (8,4) nanotube as is shown by the chiral vector in Fig.~\ref{different_cuts}(d).

The chiral vector of a nanotube is fixed by placing six pentagons onto the graphene lattice, \emph{i.e.}, the first step in the construction of the cap. This topological construction uniquely specifies the structure of the carbon cap and the tube that can be attached to it. This also determines the formation energy of the cap during nanotube nucleation and the nanotube that finally grows from the carbon nucleus. We now study how caps for different nanotube are constructed by changing the pentagon pattern.

\subsection{Constructing caps for different tube chiralities}\label{sec_different_tubes}

\begin{figure}[b]
\epsfig{file=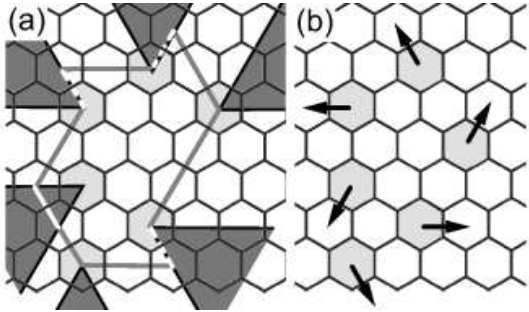,width=8.5cm,clip=}
\caption[]{(Color online) (a) (9,3) cap obtained from the (9,2) pattern in Fig.~\ref{how_to} by moving the rightmost hexagon along $\bm{a}_1$. For the cap $\bm{a}_1$ is parallel to the line of the cap hexagon (full, red). The displacement of one of the six hexagons in (b) along $\bm{a}_1$ (arrows) creates six distinct (9,3) caps (one violates the isolated pentagon rule).}
\label{move}
\end{figure}

To create different caps from a starting pentagon pattern we need to move one (or more) of the pentagons. We now show that this creates $(n_1',n_2')$ nanotubes in a rational manner. In Fig.~\ref{how_to}(a) the lines forming the cap hexagon (full, red) are parallel to the $\bm{a}_1$ direction of what later becomes the rolled up graphene sheet. Moving one pentagon along $\bm{a}_1$ we increase the length of the cap hexagon by $\bm{a}_1$ and introduce an additional $-\bm{a}_1+\bm{a}_2$ step, Fig.~\ref{move}(a). Thus, the resulting nanotube is given by $(n_1',n_2')=(n_1,n_2+1)$. In our example this is the (9,3) nanotube. For the cap the $\bm{a}_1$ direction is fixed with respect to the cap hexagon, not the graphene sheet.  At every point of the lattice in Fig.~\ref{move}(a) $\bm{a}_1$ is parallel to the line defining the cap hexagon. When the direction of the full line changes by 120$^\circ$ $\bm{a}_1$ changes as well. Therefore, the displacement of a pentagon in the directions  shown in Fig.~\ref{move}(b) also creates caps for the (9,3) nanotube. These are six out of 364 patterns that match the (9,3) nanotube or five out of 33 if we impose the isolated pentagon rule.\cite{brinkmann99} Moving the pentagon along $\bm{a}_2$ results in an additional $-\bm{a}_1+\bm{a}_2$ step, but does not change the length of the hexagonal line. We thus obtain an $(n_1-1,n_2+1)=(8,3)$ cap. All other displacements can be described as the sum of $\bm{a}_1$ and $\bm{a}_2$ displacements. 

\begin{figure}
\epsfig{file=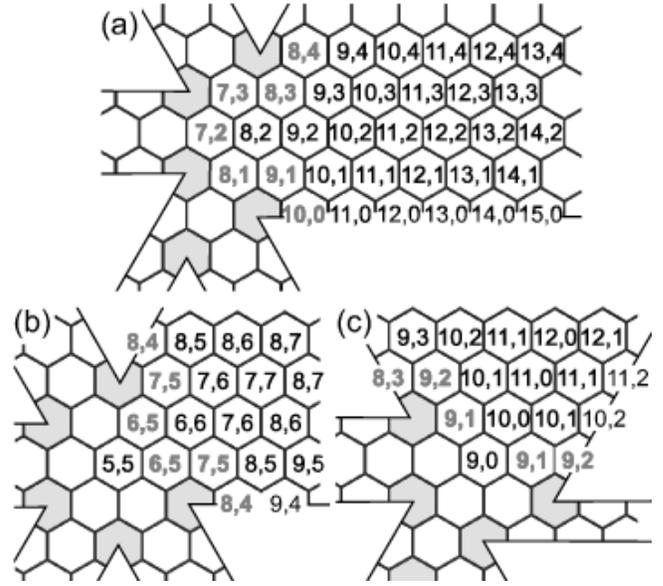,width=8.5cm,clip=}
\caption[]{(Color online) (a) Caps created from the (9,2) tube by fixing five pentagons (light grey, orange) and moving the remaining pentagon over the hexagonal lattice; (b) same as (a) but starting from the high-symmetry (5,5) configuration in Fig.~\ref{cap_55}; (c) same as (a) staring from a high-symmetry (9,0) cap. If we move the free pentagon next to another pentagon the cap violates the isolated pentagon rule (bold, red indices).}
\label{pattern}
\end{figure}

Figure~\ref{pattern}(a) shows the caps we obtain starting from the (9,2). The displacement of just one pentagon creates a large variety of chiral indices. Some of the caps in Fig.~\ref{pattern}(a) are irregular. By this we mean that five pentagons are concentrated in one part of the hexagonal lattice, whereas the sixth pentagon is far away from the others. More regular caps can be constructed by displacing more than one pentagon. Take, for example, the (12,2) nanotube in Fig.~\ref{pattern}(a). To construct a regular cap for this tube we start from the (9,2) pattern in Fig.~\ref{how_to}(a). We take the rightmost pentagon as the first pentagon [the one we moved around in Fig.~\ref{pattern}(a)] and then go counterclockwise through the six pentagons II, III, IV and so forth. Moving one after the other yields the series
\begin{multline}
(9,2)\overset{0}{\underset{\mathrm{I}}\longrightarrow}
(9,2)\overset{-\bm{a}_2}{\underset{\mathrm{II}}\longrightarrow}
(10,1)\overset{\bm{a}_1}{\underset{\mathrm{III}}\longrightarrow}
(10,2)\overset{\bm{a}_1-\bm{a}_2}{\underset{\mathrm{IV}}\longrightarrow}\\\longrightarrow
(11,2)\overset{-\bm{a}_2}{\underset{\mathrm{V}}\longrightarrow}
(12,1)\overset{\bm{a}_1}{\underset{\mathrm{VI}}\longrightarrow}(12,2),
\end{multline}
where the vectors above the arrows indicate the displacement; the roman numerals count the hexagons. The resulting (12,2) cap consists of two columns of pentagons along the armchair direction of graphene. It looks similar to the (8,2) pattern in Fig.~\ref{pattern}(a), but with more space between the armchair columns.

In Fig.~\ref{pattern}(b) and (c) we present the caps obtained starting from the (5,5) and (9,0) caps and displacing one pentagon. The high symmetry of the original pentagon patterns is reflected in the chiralities that result from the displacement of one pentagon. The patterns of five pentagons shown in Fig.~\ref{pattern}(b) and (c) in grey (orange) have mirror planes. We thus obtain pentagon patterns that are mirror images of each other when we move the remaining pentagon over the hexagonal lattice. These mirror images are (identical) caps for the same pair of chiral indices. To highlight this, we always brought the $(n_1,n_2)$ indices back into the standard $0^\circ\le\varTheta\le30^\circ$ graphene segment ($n_1,n_2>0, n_1>n_2$) in Fig.~\ref{pattern}(b) and (c). For example, the series (9,1), (10,0), (10,1) and (10,2) in Fig.~\ref{pattern}(c) is equivalent to (9,1), (10,0), (11,-1), and (12,-2), because the $(n_1,n_2)$ and $(n_1+n_2,-n_2)$ tube are identical. For tubes close to the armchair direction we interchanged $n_1$ and $n_2$ in the left part of Fig.~\ref{pattern}(b). This turns a left-handed tube into a right-handed tube; otherwise the two tubes are identical.

Many chiral indices appear in more than one panel of Fig.~\ref{pattern}. \emph{E.g.}, the (8,4) and (9,4) tubes are present in (a) and (b) and all chiral indices of (c) can be found in (a) as well. Some of the repeated indices correspond to the same pentagon pattern, \emph{i.e.}, the same cap, see the (10,1) caps in Fig.~\ref{pattern}(a) and (c). Most repeated indices, however, describe two distinct caps for one nanotube chirality. For example, the (10,0) cap in Fig.~\ref{pattern}(c) is very regular and obeys the isolated pentagon rule, whereas the (10,0) cap in (a) contains two adjacent pentagons.

\section{Formation energies of (10,0) caps}\label{sec_formation}

Our original motivation for studying nanotube caps was to calculate the formation energies of the caps on catalytic particles.\cite{reich05c} We found that the two important contributions for the total energy of a nanotube nucleus were the cap formation energy and the carbon-metal binding energy. Here the question arises whether distinct caps matching the same nanotube have different formation energies and would hence be preferred in a growth process. To answer this question we calculated the total energies of (10,0) caps from first-principles.

\begin{table*}
\caption[]{Formation energy of (10,0) caps. $E_c$ is the formation energy of the cap alone; $E_{60}$ corresponds to the formation energy for a half capped tube containing 60 carbon atoms, see text for details. The formation energy for the (10,0) tube and $C_{70}$ are given for comparison. All energies are referred to the total energy of a graphene sheet. ``$-$'' means no structure exists with isolated pentagons or this number of adjacent pentagons for a given $n_a$.}
\label{tab_formation}
\begin{ruledtabular}
\begin{tabular}{lD{.}{.}{2}D{.}{.}{4}D{.}{.}{4}D{.}{.}{2}D{.}{.}{4}D{.}{.}{4}D{.}{.}{2}D{.}{.}{4}D{.}{.}{4}}
$n_a$&\multicolumn{3}{c}{isolated}&\multicolumn{3}{c}{two adjacent}&\multicolumn{3}{c}{four adjacent}\\
&\multicolumn{1}{c}{$E_c$ (eV)}&\multicolumn{1}{c}{$E_c$/C (eV)}&\multicolumn{1}{c}{$E_{60}$ (eV)}&
\multicolumn{1}{c}{$E_c$ (eV)}&\multicolumn{1}{c}{$E_c$/C (eV)}&\multicolumn{1}{c}{$E_{60}$ (eV)}&
\multicolumn{1}{c}{$E_c$ (eV)}&\multicolumn{1}{c}{$E_c$/C (eV)}&\multicolumn{1}{c}{$E_{60}$ (eV)}\\[0.5ex]\hline
$40$	&14.8	&0.370&17.5	&-&-&-		&17.2&0.438&20.0\\
$42$	&14.4&0.342&16.8	&15.7&0.375&18.2	&16.7&0.400&19.2\\
		&14.7&0.350&17.2\\
$46$	&16.0&0.348&17.9	&17.3&0.376&19.2	&19.4&0.422&21.3\\
$48$	&15.3&0.320&17.0	&17.3&0.359&18.9	&20.2&0.421&21.8\\
$52$	&17.0&0.328&18.1	&17.1&0.329&18.2	&19.1&0.367&20.2\\
$54$	&-&-&-		&18.9&0.350&19.7	&21.2&0.392&22.0\\
$60$	&17.4&0.290&17.4&18.8&0.313&18.8	&19.6&0.326&19.6\\[0.5ex]\hline
average	&&&17.4&&&18.8&&&20.6\\[0.5ex]\hline
(10,0) tube&&0.137&8.2\footnote{formation energy of 60 atoms of an infinite (10,0) nanotube}\\
half $C_{70}$&13.2&0.376&17.8\footnote{Calculated using 40 $C_{70}$ atoms for the cap and 20 (10,0) atoms for the tube segment. This segment is then equivalent to the 40 atoms cap in the first row.}\\
\end{tabular}
\end{ruledtabular}
\end{table*}

\emph{Ab-initio} calculations were performed using the \textsc{Siesta} code.\cite{soler02} The core electrons were described by non-local norm-conserving pseudopotentials\cite{troullier91}, the valence electrons by a double-$\zeta$ basis set.\cite{junquera01} The cutoff radii were $4.2\,$a.u. for the $s$ and $5.0\,$a.u. for the $p$ orbitals. The cutoff in real space was $\approx300\,$Ry. We used the generalized gradient approximation (GGA) as parameterized by Perdew, Burke, and Ernzehof.\cite{perdew96} These input parameters are the same as we used for our calculations of the nanotube caps on Ni.\cite{reich05c} The starting cap structures were obtained using the \textsc{CaGe} program.\cite{brinkmann99} Two caps were joined to form a fullerene and placed into a cubic unit cell. The cell length was 20\,{\AA}, \emph{i.e.}, interactions between repeated images were strictly zero because of the finite length of the basis functions.\cite{soler02} The fullerenes were relaxed by a conjugate gradient optimization until all forces were below $0.04\,$eV/{\AA}. Our formation energies are given with respect to the total energy of a graphene sheet with the same number of carbon atoms.

The (10,0) tube has seven caps obeying the isolated pentagon rule.\cite{brinkmann99,astakhova99} We calculated the formation energy of all these seven caps plus 13 caps with adjacent pentagons (out of 251). 
The formation energies we obtained are given in Table~\ref{tab_formation}. The caps contained between 40 and 60 carbon atoms. For a fixed number of adjacent pentagons the formation energy $E_c$ per carbon atom decreases with increasing number of atoms in the cap. This was expected, because the formation energy of fullerenes scales with $\ln{N_6}$, where $N_6$ is the number of hexagons in the fullerene.\cite{tersoff92,zhang92} We corrected for this dependence by adding the energy of carbon atoms in the (10,0) nanotube $E_c(10,0)$ to obtain a constant number of hexagons (or carbon atoms). This is equivalent to assuming that the chemical potential for carbon is controlled by the sides of the tube. $E_{60}$ in Table~\ref{tab_formation} thus represents a capped (10,0) segment with 60 atoms 
\begin{equation}
E_{60}=E_c(\mathrm{cap}) + (60 - n_a) E_c(10,0),
\end{equation}
where $E_c(\mathrm{cap})$ is the \emph{ab-initio} formation energy for the cap, $n_a$ the number of carbon atoms in the cap and $E_c(10,0)=0.137\,$eV/C, see Table~\ref{tab_formation}.

The average formation energy for a segment with 60 carbon atoms is $17.4\,$eV if the cap obeys the isolated pentagon rule. $E_{60}$ varies by up to $\pm0.7\,$eV for the different caps; the most stable structure is half a $C_{84}$ fullerene with $E_{60}=16.8\,$eV. Allowing one pair of adjacent pentagons $E_{60}$ increases by $1.4\,$eV or 8\%. A notable exception is the cap with 52 atoms that only differs by 0.1\,eV between the isolated-pentagon and two adjacent-pentagons cap. With two pentagon pairs the average formation energy is $3.2\,$eV (18\%) higher than for isolated pentagons. The typical formation energy of a pair of adjacent pentagons is thus around $1.5\,$eV. Note that this is $\approx1/4$th of the energy necessary for a Stone-Wales defect ($5.3-6$\,eV, Refs.~\onlinecite{jensen02,li05}).

\begin{figure}[b]
\epsfig{file=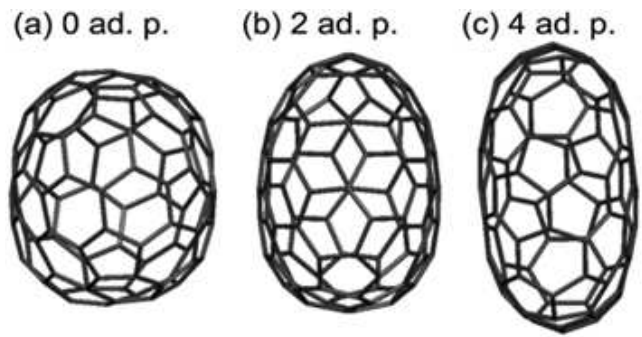,width=8.5cm,clip=}
\caption[]{(Color online) Relaxed fullerene equivalent of the nanotube caps with 48 carbon atoms and (a) isolated pentagons, (b) two adjacent pentagons and (c) four adjacent pentagons, compare Table~\ref{tab_formation}. The $z$ axis of an attached (10,0) tube should be thought of as perpendicular to the paper.}
\label{cap_96}
\end{figure}

The increase in formation energy as shown in Table~\ref{tab_formation} is, in fact, only a lower boundary for the energy required for caps with adjacent pentagon pairs. The reason for this is the geometry of the relaxed fullerenes with zero, two and four pentagon pairs. The relaxed fullerenes obeying the isolated pentagon rule typically had a circular cross section, \emph{i.e.} the part of the cap where the tube would be attached. The fullerenes constructed from caps with two or more adjacent pentagon pairs, in contrast, were elliptical. Figure~\ref{cap_96} shows as an example the fullerenes corresponding to the cap with 48 carbon atoms in Table~\ref{tab_formation}. The cross section changes from almost circular for the isolated pentagon cap in Fig.~\ref{cap_96}(a) to elliptical for the caps with two and four adjacent pentagons. The cross section in Fig.~\ref{cap_96}(c) has an aspect ratio close to 2. Attaching an elliptical cap to a circular nanotube induces additional strain in the cap or a part of the tube. This results in a further increase in the energy required for the formation of adjacent pentagons. Relaxed fullerene equivalents of caps with adjacent pentagons are flat, because the adjacent pentagon pairs create large curvature in a part of the cap. The remaining part of the cap contains mainly hexagons and becomes flat. This gives rise to the elliptical shape of the relaxed caps. 

The $\approx1.4$ and $3.2\,$eV difference in formation energy for caps with adjacent pentagons is on the same order or larger than the demarcation energy of carbon nanotubes during nucleation ($\approx 2.8\,$eV at 1000\,K, see Ref.~\onlinecite{reich05c}). By this we mean the energy difference necessary for the exclusive growth of a specific carbon structure. When the formation energy of two carbon nuclei differs by more than the demarcation energy, the nucleus with the higher energy is formed with a very small yield. We thus find that for low-temperature growth, the formation of nanotube caps obeying the isolated pentagon rule is much preferred. The (10,0) tube has caps obeying the isolated pentagon rule. Its cap and hence the (10,0) tube can grow when isolated pentagons are preferred energetically. This changes dramatically when we consider single-walled carbon nanotubes with smaller diameters. We now show that the narrow diameter distribution in certain CVD grown nanotubes can be understood from the formation energy of adjacent pentagon pairs.

\section{Adjacent pentagons and diameter distribution of CVD samples}\label{sec_abundance}

\begin{figure}
\epsfig{file=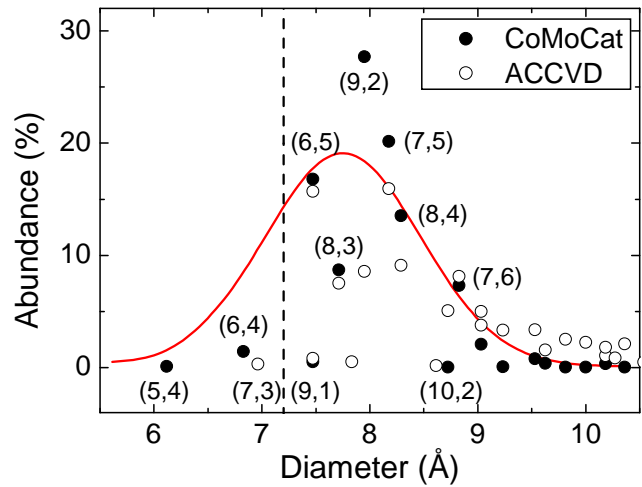,width=8.5cm,clip=}
\caption[]{(Color online) Abundance of nanotubes as determined from photoluminescence (PL) spectroscopy \emph{versus} tube diameter. PL data were taken from Bachilo~\emph{et al.}\cite{bachilo03} (CoMoCat, closed symbols) and Miyauchi~\emph{et al.}\cite{miyauchi04} (ACCVD, open) and normalized to the intrinsic tube intensities following Reich~\emph{et al.}\cite{reich05b}. Tubes with diameters smaller than 7.25\,{\AA} have only caps that violate the isolated pentagon rule (dashed line). The full line is a guide to the eye (Gaussian centered at 7.75\,{\AA} with $\sigma=1.7\,${\AA}). The chiral indices are indicated for selected nanotubes. Only semiconducting tubes can be detected by PL.}
\label{abundance}
\end{figure}

Bachilo~\emph{et al.}\cite{bachilo03} and Miyauchi~\emph{et al.}\cite{miyauchi04} reported single-walled nanotube samples with mean diameters below 10\,{\AA}. The remarkable thing about these tubes was that they showed a very narrow diameter and chiral angle distribution. To measure the chirality distribution both groups used photoluminescence.\cite{bachilo02,bachilo03,miyauchi04} This has to be treated with some care, because the luminescence cross section varies from tube to tube.\cite{reich05b} This is most important for comparing tubes with different chiral angles. In our analysis here we will concentrate on the diameter dependence of the tube abundance.

In Fig.~\ref{abundance} we show the abundance of tubes reported by Bachilo~\emph{et al.}\cite{bachilo03} and Miyauchi~\emph{et al.}\cite{miyauchi04}. The data were corrected by the calculated photoluminescence intensities obtained by us.\cite{reich05b} The correction depends on diameter as $1/d^2$, and in a non-trivial way on the chiral angle and nanotube index family.\cite{reich05b,reich00c} The conclusions presented here are, however, insensitive to the details of the correction.

Starting from large diameters, the abundance of nanotubes in the CoMoCat (Ref.~\onlinecite{bachilo03}) and the ACCVD (Ref.~\onlinecite{miyauchi04}) sample in Fig.~\ref{abundance} increases with decreasing diameter. It peaks around $7.8\,${\AA}. Below 7.2\,{\AA} the abundance of tubes drops sharply to zero. This drop happens exactly at the radius that separates semiconducting nanotubes into tubes with isolated pentagon caps [(6,5), (9,1), (7,5), and larger] and tubes that only have caps with adjacent pentagons [(6,4), (7,3), and smaller], see the chiral indices and the dashed line in Fig.~\ref{abundance}. The thermal energy at nucleation was thus too small for the formation of adjacent pentagons, which cost 1.5\,eV per pair as we showed above.

The formation energy of the tubes with smallest diameters increases further when taking into account the curvature energy of the tube in addition to the energy required for adjacent pentagons. The (6,4) cap, for example, contains 30 carbon atoms. An adjacent pentagon pair thus costs $\approx50\,$meV per atom in the cap. The difference in strain energy between the (6,4) and the (6,5) nanotube is $\approx30\,$meV/C (Ref.~\onlinecite{sanchez99a}). The total energy difference between a capped (6,4) and (6,5) nanotube (neglecting the catalyst) will thus decrease from $80\,$meV/C for the cap alone to 30\,meV/C for a long tube where the effect of the adjacent pentagon becomes negligible. From \emph{ab-initio} calculations we found a difference in formation energy 100\,meV/C for the (6,4) and the (6,5) cap in good agreement with the estimate given above.\cite{reich05c} It is, however, important to realize that the dependence of the strain energy on tube diameter will never explain, \emph{e.g.}, the low abundance of the (9,1) nanotube as compared to the (6,5), see Fig.~\ref{abundance}. These two tubes have exactly the same diameter. The formation of the nanotube nucleus---the cap on the metal particle---is a limiting step for the growth of a tube. 

The line in Fig.~\ref{abundance} is a guide to the eye. It is a Gaussian with a mean diameter $d\approx7.75\,${\AA} and $\sigma=1.7\,${\AA}. As can be seen, the abundance of tubes follows reasonably well a Gaussian distribution for large diameters, but the tail towards small diameters is missing. In particular, there is a marked asymmetry between the very small or vanishing abundance below 7.2\,{\AA} and the comparatively broad distribution towards the large-diameter end in Fig.~\ref{abundance}. 

The low-temperature CVD experiments confirm the importance of the cap structure and the cap formation energy for the distribution of nanotube chiralities in a sample.\cite{reich05c} Once a cap is formed, it determines the $(n_1,n_2)$ nanotube that grows from it. This is somewhat similar to the growth of tubes using other nanotubes as a seed.\cite{wang05b} If the formation energy of a certain cap is larger than the energy available during nucleation, this prevents the growth of the tube corresponding to the cap. 
This holds even when the tube itself is otherwise favourable energetically.

In this study we only considered the energy of the cap itself. The second important contribution to the formation energy during nucleation is the carbon-metal binding energy.\cite{reich05c} This can, in particular, introduce energy differences between caps of similar diameter and hence similar cap formation energy (curvature energy). The variations in the total carbon-metal binding energy ($1-2$\,eV) are similar to the energy required for adjacent pentagons (1.5\,eV). The carbon-metal binding energy can thus be the origin of the preferential growth of certain chiralities, whereas the adjacent-pentagon energy prevents the growth of tubes with very small diameters.

\section{Conclusions}\label{sec_conclusion}

In summary, we studied the structure and energetics of nanotube caps. We showed that the arrangement of pentagons in the cap defines the chirality of the tube that matches to it. Moving one (or more) pentagons across the hexagonal lattice creates caps for different nanotubes in a rational way. The isolated-pentagon caps for a (10,0) nanotube vary in formation energy by $\pm0.7\,$eV or $12\,$meV/atom. Introducing adjacent pentagons requires an energy of $\approx1.5\,$eV per pentagon pair. The large formation energy for two adjacent pentagons explains why tubes with diameters below $7.2\,${\AA} had a very small yield in low-temperature CVD growth. Our study shows that the structures and energetics of carbon caps on a catalytic particle will be the key for chirality selective growth of single-walled carbon nanotubes. A nanotube cap on a catalytic particle is the nucleus of a tube and uniquely determines its chirality.

\acknowledgements

This work was supported by the EC project CARDECOM. S. R. acknowledges financial support by the Oppenheimer Trust and Newnham College, Cambridge, UK.

%\bibliographystyle{apsrev}
%\bibliography%{d:/paper05/lss/biblio/buch,d:/paper05/lss/biblio/buch_jm,d:/paper05/lss/biblio/str_buch,d:/paper05/lss/biblio/bib_CT,d:/paper05/l%ss/biblio/lss,d:/paper05/lss/biblio/lss_str,d:/paper05/nucleation/paper/prl_resub/growth}

\end{document}